\documentclass{iaus}
\usepackage{graphicx}

\title[~~Multi-component parametric inversion] 
{Multi-component parametric inversion of galaxy kinematics and stellar populations using full spectral fitting.}

\author[Katkov \& Chilingarian]   
{Ivan Yu. Katkov$^1$
 \and Igor V. Chilingarian$^1$}

\affiliation{$^1$Sternberg Astronomical Institute, Moscow State University, \\ 
Universitetskii pr. 13, Moscow, 119992 Russia 
\\ email {\tt IK: katkov.ivan@gmail.com; IC: chil@sai.msu.ru} \\
[\affilskip]}

\pubyear{2011} 
\volume{284}  
\pagerange{1--12}
\jname{The Spectral Energy Distribution of Galaxies}
\editors{R.J. Tuffs \&  C.C.Popescu, eds.}
\begin{document}

\maketitle

\begin{abstract} 
The stellar
line-of-sight velocity distribution (LOSVD) can be strongly asymmetric in
regions where the light contributions of both disc and bulge in spiral and
lenticular galaxies are comparable.
Existing techniques for the stellar kinematics analysis do not take into
account the difference of disc and bulge stellar populations. Here we
present a novel approach to the analysis of stellar kinematics and stellar
populations. We use a two-component model of spectra where different stellar
population components are convolved with pure Gaussian LOSVDs. For this
model we present Monte-Carlo simulations demonstrating degeneracies between
the parameters.

\end{abstract}

\firstsection 
\section{Introduction}

A flat rotating stellar disc and a slowly rotating spheroidal bulge in
spiral and lenticular galaxies usually possess very different stellar
population properties. Consequently, the resulting stellar line-of-sight
velocity distribution (LOSVD) can be strongly asymmetric in regions where
the light contributions of both disc and bulge are comparable. At the first
approximation, this can be accounted by the Gauss-Hermite parametrization of
the LOSVD (\cite[van der Marel \& Franx, 1993]{vdMF93}), however, different
absorption features (e.g. sensitive to age and metallicity) will have
different effective LOSVDs. The first attempt to recover parametrically the
multi-component dynamics connected to multi-component stellar populations
was done by \cite[De Bryuine et al. 2004]{DBDRDZ04} where the authors used
stellar spectra to model different stellar populations. Here we present a
more realistic approach based on the full spectral fitting, where multiple
dynamical components are represented by multiple simple stellar population
(SSP) models. More sophisticated stellar population models can be used
instead of SSPs.

\section{Two-component parametric LOSVD recovery} 

As an example, here we present a two-component model having two pure
Gaussian LOSVD components with different stellar populations characterised
by their ages and metallicities. An optimal template is represented by the
linear combination of two SSPs each convolved with its own LOSVD, hence the
$\chi^2$ value is computed as follows:

\begin{equation}
 \chi^2=\sum_{N_\lambda} \frac{[ F_i - P_p \cdot \sum_{j} k_j \cdot S(T_j,Z_j) \otimes \mathcal{L}(v_j,\sigma_j)]^2}{\delta F_i^2},
\end{equation}
where $\mathcal{L}(v,\sigma)$ - pure Gaussian LOSVD; $F_i$ and $\delta F_i$
are observed flux and its uncertainty; $S(T_j,Z_j)$ is the flux from the
$j$-th synthetic spectrum of SSP with given age $T_j$ and metallicity $Z_j$;
$P_p$ is multiplicative Legendre polynomials of order $p$ for correcting the
continuum which determined at each step of minimization loop by solving the
linear least-square problem; $k_j$ is the j-th component weight (normally
found by the linear minimization). The important point in this study is that we fixed the relative SSP
contributions $k_j$ to the values derived from the photometric light profile
decomposition. This approach was implemented on top of the \textsc{NBursts}
full spectral fitting technique (\cite{Chil07a,Chil07b}).

Our new approach was applied to the real spectra of luminous early-type
galaxy NGC~524 (see \cite[Katkov et al. (2011)]{Katkov11}). It was shown
that LOSVD of NGC~524 exhibits strong asymmetry and a stellar component of
galaxy can be described as a bulge, inner disc and a conterrotating outer
disc.

\section{Monte-Carlo simulation} 

\begin{figure}
\begin{center}
 \includegraphics[width=0.95\textwidth]{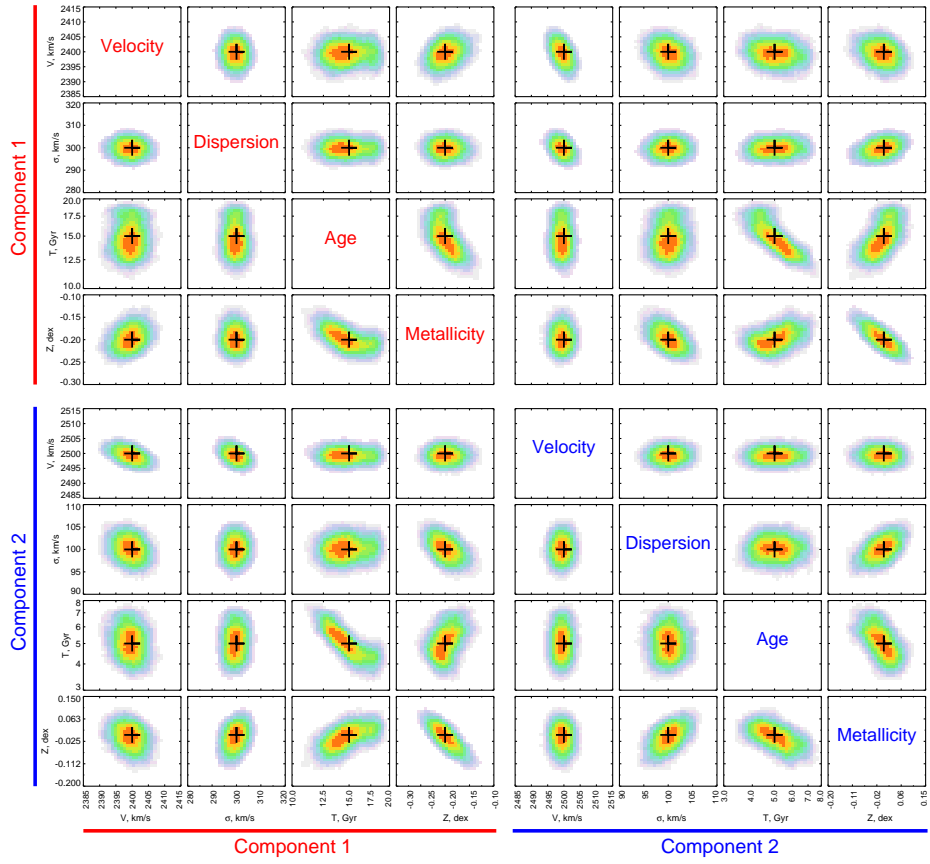} 
 \caption{Covariance matrix of the model parameters. Left-top quadrant and
 right-bottom quadrant present a self-correlation of parameters of bulge and
 disc components, correspondingly. Left-bottom and right-top quadrants
 correspond to mutual correlation of stellar components. White crosses show
 the input values.}
   \label{fig_mc}
\end{center}
\end{figure}

In order to explore the parameter space, we constructed and analysed using
our technique a sample of 8000 realizations of synthetical spectra created
as a linear combination of two components (see Table~\ref{tabl_parameters})
adding the random noise corresponding to the signal-noise ratio (SNR) 100.
The relative contributions $k_{1,2}$ were fixed. The degeneracies between
all the parameters are shown in Fig.~\ref{fig_mc} as a covariance matrix.
Here and after red and blue points correspond to the bulge and disc
components respectively. We can see that the most degenerated pairs of
parameters are $T_1-T_2$ and $Z_1-Z_2$.

\begin{table}
  \begin{center}
  \caption{Parameter of stellar components.}
  \label{tabl_parameters}  
  \begin{tabular}{lccccc}
  \hline
  Component & $v$, km/s & $\sigma$, km/s & Age, Gyr & Z, dex & Weight $K_j$\\
  \hline
   1 (Bulge) & 0 & 300 & 15 & -0.2 & 0.7\\
   2 (Disc) & 100 & 100 & 5 &  0.0 & 0.3\\
  \hline
  \end{tabular}
 \end{center}
\end{table}

\begin{figure}
\begin{center}
 \includegraphics[width=0.95\textwidth]{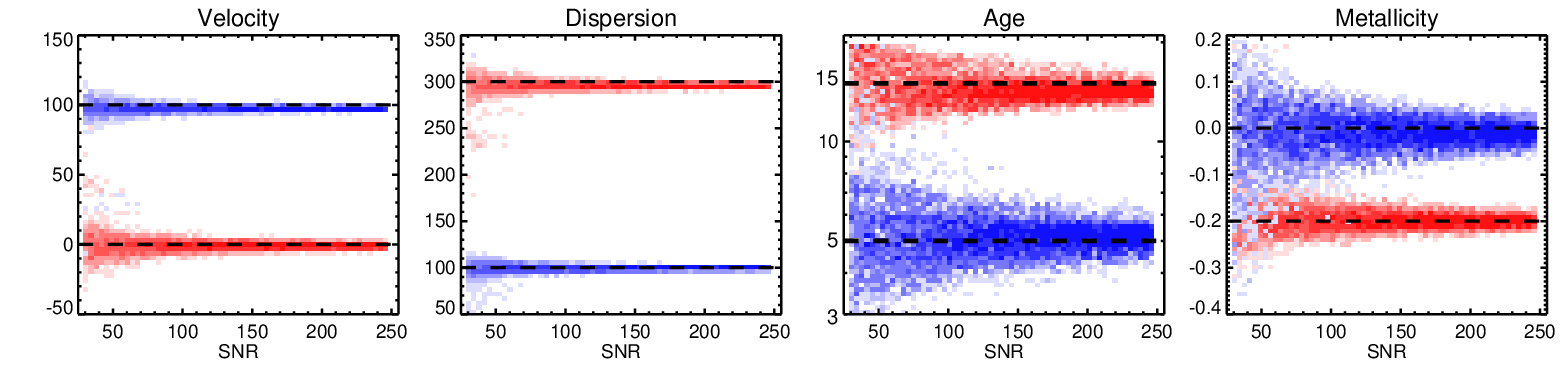} \\
  \includegraphics[width=0.95\textwidth]{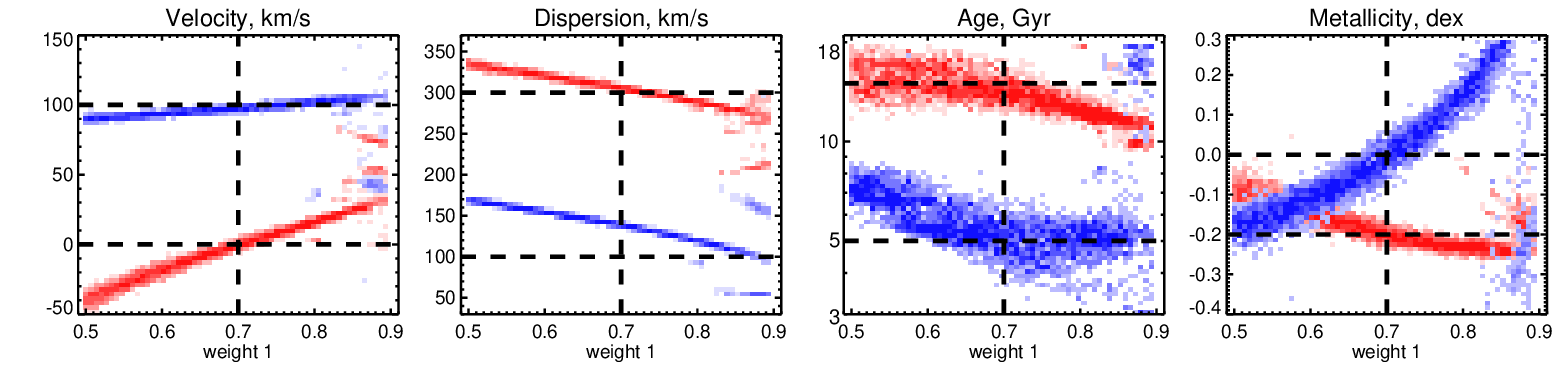}
 \caption{Top line: output parameter dispersion as a dependence on signal-to-noise ratio. Bottom line: output parameters with varied relative contribution of components. }
   \label{fig_snr}
\end{center}
\end{figure}

In order to demonstrate how the parameters are recovered depending on the
SNR we fitted a sample of 8000 models varying a SNR between 30 and 250.
Fig.~\ref{fig_snr}(top line) shows how the output parameter uncertainty
decreases when increasing SNR. The dashed lines show the input parameters of
synthetic spectrum.

We used the same approach to demonstrate how important is the knowledge of
relative contributions $k_j$. Fig.~\ref{fig_snr}(bottom line) presents 4000
MC realizations where we varied the relative SSP contribution of bulge
component $k_1$ in range between 0.5 and 0.9 while the real value was equal
to 0.7. SNR=150 was adopted in these realisations. It is obvious that even
little offset in $k_j$ can lead to important biases in the recovered
parameters.

\bigskip

Authors thank the IAU for the provided financial aid and RFBR grant 10-02-00062а for covering the remaining expanses.

\end{document}